# Comment on "The Jones-Hore theory of radical-ion-pair reactions is not self-consistent" (arXiv:1010.3888v3)

Kiminori Maeda

*Department of Chemistry, University of Oxford, Centre for Advanced Electron Spin Resonance (CAESR), Inorganic Chemistry Laboratory, Oxford OX1 3QR, U. K.*

Prompted by Kominis's formulation of a theory based on quantum measurement concepts [1], there has been a debate on how the spin-selective chemical reactions of a radical pair (RP) should be described in the equation of motion of the spin-density matrix. To date, there are three models: (1) The conventional model proposed by Haberkorn in 1976 [2], (2) A quantum Zeno approach by Kominis [1], and (3) the Jones-Hore model [3]. The Haberkorn model was proposed with no reference to quantum measurement concepts. It has been well accepted [4-6] and used extensively in the field of Spin Chemistry for many years. Although the Jones-Hore form of the recombination superoperator was formulated using quantum measurement arguments, the difference between it and the Haberkorn form is slight. In contrast, Kominis's model predicts very different spin dynamics from the other two and could therefore have serious consequences for the analysis of RP spin dynamics.

In Ref. [7], Kominis claims to have found an inconsistency in the Jones-Hore model of spin-selective RP recombination [7]. The essence of this claim is as follows. We consider the simplest case, in which only the singlet RP state can recombine, and ignore triplet recombination and the coherent spin dynamics arising from the spin Hamiltonian. According to the Jones-Hore approach [3], the time evolution of the (unnormalized) RP density matrix $\rho$ is written

$$\frac{d\rho}{dt} = -k_S \left( \rho - Q_T \rho Q_T \right) \qquad \text{(eq.1 in ref. [7])} \qquad (1)$$

Where $k_S$ is the recombination rate constant from the singlet state $|S\rangle$ of the RP and $Q_T$ is the projection operator onto the triplet state $|T\rangle$. Now we define the normalized density matrix of the non-reacting RP by

$$\rho_{nr} = \rho / \text{Tr}\{\rho\} \qquad \text{(eq. in ref. [7])} \qquad (2)$$

so that:

$$\frac{d\rho_{nr}}{dt} = -k_S \text{Tr}\{Q_T \rho_{nr} Q_T\} \left[ \rho_{nr} - \frac{Q_T \rho_{nr} Q_T}{\text{Tr}\{Q_T \rho_{nr} Q_T\}} \right] \qquad \text{(eq.2 in ref. [7])} \qquad (3)$$

Kominis has also used an alternative route to $d\rho_{nr}/dt$ and obtained a different result:

$$\frac{d\rho_{nr}}{dt} = -k_S \left[ \rho_{nr} - \frac{Q_T \rho_{nr} Q_T}{\mathrm{Tr}\{Q_T \rho_{nr} Q_T\}} \right] \qquad \text{(eq.3 in ref. [7])} \qquad (4)$$

Based on the discrepancy between eqs. (3) and (4), Kominis has claimed that the Jones-Hore theory is inconsistent. However, this formulation includes a misconception about the Jones-Hore approach. In the present comment, I re-formulate the derivation of $d\rho_{nr}/dt$, correcting the error in Kominis's treatment. In his paper [7], Kominis considers the time evolution of $\rho_{nr}$ as a kinetic process from the unmeasured RP $\rho_0$ to the measured* (projected into the triplet) RP $\rho_T$ such that

$$\rho_{nr} = \omega_0 \rho_0 + \omega_T \rho_T \qquad \text{(eq. in ref. [7])} \qquad (5)$$

where $\omega_0$ and $\omega_T$ are

$$\omega_0 = \exp(-k_S t) \quad \text{and} \quad \omega_T = 1 - \exp(-k_S t) \qquad \text{(eq. in ref. [7])} \qquad (6)$$

However, eq. (6) cannot be accepted if we interpret Jones and Hore correctly. Converting the recombination process of the Jones-Hore approach into Kominis's picture, we obtain the scheme:

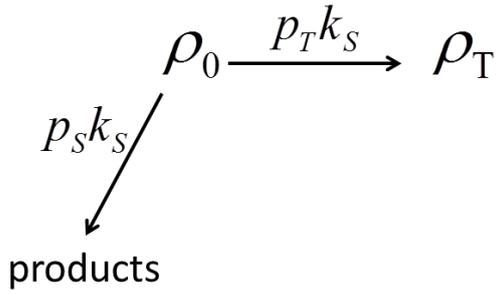

This scheme simply indicates that the singlet fraction $p_S (= \mathrm{Tr}(Q_S \rho_0))$ of $\rho_0$ recombines to the product while the triplet fraction $p_T (= \mathrm{Tr}(Q_T \rho_0))$ of $\rho_0$ changes into $\rho_T$ by a projection of the wave function onto the triplet spin state. Now we put $f_0$ and $f_T$ as the probability of $\rho_0$ and $\rho_T$, respectively. The rate equations for $f_0$ and $f_T$ are

$$\frac{df_0}{dt} = -k_S(p_S + p_T)f_0 = -k_S f_0$$
$$\frac{df_T}{dt} = p_T k_S f_0 \qquad (7)$$

which have the solutions

$$f_0 = \exp(-k_S t)$$
$$f_T = p_T \{1 - \exp(-k_S t)\} \qquad (8)$$

The factor $p_T$ in eq. (8) is different from Kominis's $\omega_0$ and $\omega_T$ in eq. (6). Of course the probability $f_0 + f_T$ is less than 1 because we have not included the probability of product formation, $f_P$. However, we can normalize $\rho_{nr}$ by rewriting $\omega_0$ and $\omega_T$ in eq. (5) as follows

$$\omega_0 = \frac{f_0}{f_0 + f_T}, \quad \omega_T = \frac{f_T}{f_0 + f_T} \tag{9}$$

Now we derive $d\rho_{nr}/dt$. First,

$$\frac{d\omega_0}{dt} = -\frac{d\omega_T}{dt} = -k_S \omega_0 (\omega_T + p_T \omega_0) = -k_S \omega_0 \mathrm{Tr}[Q_T \rho_{nr} Q_T]$$

$$\because \frac{df_0}{dt} = -k_S f_0, \quad \frac{df_T}{dt} = k_S P_T f_0 \text{ and } Q_T \rho_{nr} Q_T = (\omega_T + p_T \omega_0)\rho_T \tag{10}$$

Second

$$\rho_T = \frac{Q_T \rho_{nr} Q_T}{\mathrm{Tr}[Q_T \rho_{nr} Q_T]}, \quad \rho_0 = \frac{1}{\omega_0}\left(\rho_{nr} - \omega_T \frac{Q_T \rho_{nr} Q_T}{\mathrm{Tr}[Q_T \rho_{nr} Q_T]}\right) \tag{11}$$

And finally, using eqs (10) and (11), we obtain

$$\frac{d\rho_{nr}}{dt} = \frac{d\omega_0}{dt}\rho_0 + \frac{d\omega_T}{dt}\rho_T \tag{12}$$

which is identical to eq. (3). We conclude therefore that the Jones-Hore approach [3] is consistent at least on this point.

**Footnote**

*If a measurement target the singlet RP, the rest of the probability is in the triplet. Therefore, the failure of the measurement of singlet RP causes the projection of the wave function into the triplet RP. It can be called a null measurement [3,8].